\documentclass{ws-procs975x65}

\newcommand{\CS}{{\mbox{\tiny CS}}}

\begin{document}

\title{Towards Tests of Alternative Theories of Gravity with LISA}

\author{Carlos~F.~Sopuerta$^{1}$ and Nicol\'as~Yunes$^{2}$}

\address{$^{1}$Institut de Ci\`encies de l'Espai (CSIC-IEEC), \\
Facultat de Ci\`encies, Campus UAB, Torre C5 parells, 
Bellaterra, 08193 Barcelona, Spain\\
$^{2}$Physics Department, Princeton University, Princeton, NJ 08544, USA}

\begin{abstract}
The inspiral of stellar compact objects into massive black holes, usually known 
as extreme-mass-ratio inspirals (EMRIs), is one of the most important sources of 
gravitational-waves for the future Laser Interferometer Space Antenna (LISA). Intermediate-mass-ratio inspirals (IMRIs
are also of interest to advance ground-based gravitational-wave observatories. We discuss here how 
modifications to the gravitational interaction can affect the signals emitted by these systems and their detectability by LISA.  
We concentrate in particular on Chern-Simons modified gravity, a theory that emerges 
in different quantum gravitational approaches. 
\end{abstract}


\bodymatter

\section{Motivation}
One of the primary goals of the future {\emph{Laser Interferometer Space Antenna}} (LISA) is to 
search for modifications to General Relativity (GR) in physical situations where strong gravitational
fields are involved, speeds are large and binary pulsar or Solar System experiments do
not lead to stringent constraints.
%
One avenue to study such situations is to explore deviations from 
the Kerr metric {\emph{within GR}}.  However, in GR, the Kerr metric is considered (provided cosmic 
censorship and causality hold) to be the {\emph{only}} possible final state of the gravitational collapse 
of compact bodies. 
Another route to test GR is to consider alternative theories of gravity and study the imprint these leave on 
gravitational waves (GWs) emitted by LISA sources.  Here, we report results related to the capture and 
inspiral of stellar compact objects into massive black holes (BHs) in galactic centers, usually known
as \emph{Extreme-Mass-Ratio Inspirals} (EMRIs).

Due to the plethora of alternative theories available, it is difficult to justify the choice of one theory over another. 
One approach is to propose certain criteria for a theory to be a reasonable candidate to test GR with LISA~\cite{bias}. 
An example of such a theory is Chern-Simons (CS) modified gravity~\cite{Alexander:2009tp}. 
In this 4D theory, the Einstein-Hilbert action is modified through the addition of the product
of a dynamical scalar field and the Pontryagin density.  This modification arises generically and unavoidingly in the 
low-energy effective limits of string theory and can also arise in loop quantum gravity when the Barbero-Immirzi parameter 
is promoted to a field. Regarding BH solutions, one has to distinguish between the dynamical and non-dynamical
versions of the theory. In the latter the scalar field is a given function which leads to an additional constraint which
is too restrictive, essentially disallowing spinning BH solutions for time-like scalar fields~\cite{Grumiller:2007rv} and forbidding 
perturbations of non-spinning  BHs~\cite{Yunes:2007ss}.  The dynamical theory admits BH solutions, like the Schwarzschild metric and a 
modified Kerr metric~\cite{Yunes:2009hc}, which are the theory and BH solutions that we consider here.

\section{EMRIs in CS modified Gravity}
We now summarize recent results~\cite{Sopuerta:2009iy} on the study of EMRIs in CS modified gravity in steps:

(i) \emph{MBH geometry}:
Using the small-coupling and slow-rotation approximations, the exterior, stationary and axisymmetric
gravitational field of a rotating BH in dynamical CS modified gravity, in Boyer-Lindquist type 
coordinates, is given by~\cite{Yunes:2009hc} 
$ds^{2} = ds^{2}_{\rm Kerr} + 5 \xi a/(4 r^{4}) \left[1 + 12 M/(7 r) + 27 M^{2}/(10 r^{2}) \right] \sin^{2}{\theta} dt d\phi$, 
where $ds^{2}_{\rm Kerr}$ is the line element for the Kerr metric, $M$ and $a$ are the MBH mass and spin parameter, 
and $\xi= \alpha^{2}/(\kappa \beta)$,
where $\kappa = (16 \pi G)^{-1}$ is the gravitational constant and $(\alpha,\beta)$ are the coupling constants associated
with the gravitational CS correction and the strength of the CS scalar field respectively. 
The multipolar structure of the modified metric remains completely
determined by only two moments (no-hair or two-hair theorem): the mass monopole and the current dipole. The relation, however, 
between these two moments and higher-order ones is modified from the GR expectation at multipole $\ell \geq 4$.

(ii) \emph{The CS scalar field}: It is axisymmetric and fully determined by the MBH geometry~\cite{Yunes:2009hc}.
Hence, the no-hair theorem still holds in the dynamical theory (modulo the modification of the relation aforementioned).

(iii) \emph{Motion of test particles around the MBH}:
It has been shown~\cite{Sopuerta:2009iy} that point-particles follow geodesics in this theory, as in GR.
Geodesics in the modified MBH geometry have essentially the same properties as Kerr geodesics; 
there are enough constant of motion to completely separate the Hamilton-Jacobi equations.
The geodesic equations take the following form: $\dot{x}^{\mu} = \dot{x}^{\mu}_{\rm Kerr} + \delta x^{\mu}_{\CS}$, where 
an overhead dot denotes differentiation with respect to proper time and $x^{\mu} = [t,r,\theta,\phi]$ are 
Boyer-Lindquist-type coordinates. The corrections  
$\delta x^{\mu}_{\CS} = [L, 2 E L, 0, -E] \delta g^{\CS}_{\phi}$, where $E$ and $L$ are the energy and angular
momentum of the geodesic, and
$\delta g_{\phi}^{\CS} = \xi a/(112 r^{8} f) (70 r^{2} + 120 r\,M + 189 M^{2})$, with $f = 1 - 2 M/r$.
One can see that the innermost-stable circular orbit (ISCO) location is CS shifted by~\cite{Yunes:2009hc}:
$R^{}_{\mbox{\tiny ISCO}}= 6M \mp 4\sqrt{6}a/3 - 7a^2/(18 M) \pm 77\sqrt{6}a \xi/(5184 M^{4})$,
where the upper (lower) signs correspond to co- and counter-rotating geodesics. Notice that the 
CS correction acts {\emph{against}} the spin effects.  One can also check that the three fundamental
frequencies of motion~\cite{Schmidt:2002qk} change with respect to the GR values.

(iv) \emph{Gravitational Wave emission}:  Up to date, EMRIs in CS modified gravity have been described using
the \emph{semi-relativistic} approximation~\cite{Ruffini:1981rs}, where the motion is assumed geodesic 
and GWs are assumed to propagate in flat spacetime.  Neglecting \emph{radiation reaction} effects,
the dephasing between CS and GR GWs are due to the modifications in the MBH 
geometry.  This dephasing will not prevent in principle detection of GWs from EMRIs with LISA (from \emph{short} 
periods of data $\sim 3$ weeks,  where radiation reaction effects can be neglected), 
but instead it will bias the estimation of parameters, leading to an uncontrolled systematic error.

(v) \emph{Radiation reaction effects}: Using the short-wave approximation, it has been shown~\cite{Sopuerta:2009iy}
that to the leading order the GW emission formulae are unchanged, except for subdominant energy-momentum emission 
by the scalar field. The inclusion of radiation-reaction effects, neglected in this analysis, will be crucial in the future, 
since it will lead to stronger GW modifications that will break the degeneracy between the CS coupling parameter and the 
system parameters.

(vi) \emph{Tests of CS gravity with LISA}: 
A rough estimate~\cite{Sopuerta:2009iy} of the accuracy
to which CS gravity could be constrained via a LISA observation is
$\xi^{1/4} \lesssim 10^{5}\, {\textrm{km}} \left(\delta/10^{-6}\right)^{1/4} \left(M/M_{\bullet}\right)$,
where $\delta$ is the accuracy to which $\xi$ can be measured, which depends on the integration time, 
the signal-to-noise ratio, the type of orbit considered and how much radiation-reaction affects the orbit. 
Notice that IMRIs are favored over EMRIs. This result is to be compared 
with the binary pulsar constrained $\xi^{1/4} \lesssim 10^{4} \; {\textrm{km}}$~\cite{Yunes:2009hc}. 
We then see that an IMRI with $M = 10^{3} M_{\odot}$ could place a constraint two-orders of magnitude more
stringent than the binary pulsar one. Moreover, a GW test can constrain the dynamical behavior of the theory
in the neighbourhood of BHs, which is simply not possible with neutron star observations.

(vii) \emph{Future Work}: Current efforts focus on the inclusion of radiation reaction, a key point for which 
is that, to leading order, GW emission in CS gravity is unchanged. This simplifIes the analysis, allowing for GR-like 
expressions for the rate of change of constants of motion due to GW emission.

\section*{Acknowledgments}
We acknowledge support from NSF grant PHY-0745779,
the Ram\'on y Cajal Programme (MEC, Spain), and a Marie Curie
International Reintegration Grant (MIRG-CT-2007-205005/PHY) within the
7th European Community Framework Programme, and from the contract ESP2007-61712
of MEC, Spain.


\end{document}